\documentclass[12pt]{article}
\usepackage{amsmath}

\begin{document}
\vskip 1truein
\begin{center}
{\Large {\bf A Possible Gauge Formulation for Gravity?}}
\vskip 5pt
{\bf Rolando Gaitan D.}${}^{a, }${\footnote {e-mail:
rgaitan@fisica.ciens.ucv.ve}} \\
${}^a${\it Grupo de F\'{\i}sica Te\'orica, Departamento de F\'\i sica,
Facultad de Ciencias y Tecnolog\'\i a, Universidad de Carabobo,
Apartado Postal 129 Valencia 2001, Estado Carabobo, Venezuela.}\\
\end{center}
\vskip .5truein
\begin{abstract}

A possible Yang-Mills like lagrangian formulation for gravity is 
explored. The starting
point consists on two next assumptions. First, the metric is assumed as 
a real map from
a given gauge group. Second, a gauge invariant lagrangian density is 
considered with the
condition that it is related to the Einstein one up to a bound term. We 
study a stationary
solution of the
abelian case for the spherical symmetry, which is connected to the
M\"oller's Maxwell like formulation for gravity. Finally, it is showed 
the consistence of this
formulation with the Newtonian limit.
\end{abstract}

\section{Introduction}

 From appearance of General Relativity theory, Quadratic Lagrangian 
Formulations (QLF)
for gravity in terms of Riemann-Christofell tensor have been considered
\cite{We,Pa,Ed,La,Fa,Mi,BD,St,BFFV}. Sometimes these models are called 
"Yang-Mills type formulations".
In QLF, Einstein field equations are recovered from Palatini's variational
principle \cite{St}. On the other hand, the addition of terms which are 
quadratic in
curvature to the Einstein gravitational Lagrangian yields a theory where 
the renormalization
problems are much less severe \cite{Ste}. This is similar to the 
situation in the Yang-Mills
theories which are renormalizable \cite{Sl,tH}. These results had 
motivated some authors
to propose, for the gravitational lagrangian, expressions which are 
quadratic in
curvature \cite{Y,vdH,Fair,Ad}.

In a parallel way to this current research, and immediately after the 
formulation of
Yang-Mills theory (YMT) \cite{YM}, R. Utiyama \cite{U} was the first to 
recognize
the gauge ''character'' of gravitational field.
Taking into account the protagonical role of YMT in the electro-weak model
\cite{Gl,SW,Wei},
one could think that this kind of theory is a candidate for description 
and quantization
of fundamental interactions. Thus, there is a significant reason to 
considerate
gauge formulations for gravity. Here, we are referring to constructions
where gravitation
  is described by means of agauge connection on a certain
fibre bundle.   Within the variety of these theories
\cite{K,C,Ki,A,TK,B,Bra}, it must be detached the Hamiltonian
formulation of
A. Ashtekar \cite{A} from which a program for quantization of
gravity
could be performed.

In this paper we present a possible quadratic gauge lagrangian formulation
(i.e., Yang-Mills like construction)
for gravity linear in Ricci curvature, following the YMT and in contrast 
to QLF.
The starting point consists to consider real mappings of a connection
in order to obtain a metric. This idea is not new, i.e., in \cite{Bra} 
are presented
new gauge theories of conformal space-time symmetries which merge 
features of YM theory
and gravity.

The paper is organized as follows.
In section 2, we present two primary considerations
for the construction of a gauge invariant Lagrangian formulation,
starting from the existence of a one-form connection that transforms under
a given Lie group $G$.
In section 3, field equations and the relationship with
the Einstein ones are presented.
Next, in section 4, we study the abelian theory computing a stationary 
solution for
the connection. It is shown that there is a particular solution proportional
to $U(1)$ connection of Maxwell type formulation of M\"oller for gravity.
Finally, in section 5 the consistence between the Yang-Mills like field 
equations
for gravity and Newtonian limit is discussed. We conclude with some remarks.

\section{A gauge Lagrangian formulation }

Let $G$ be a Lie group, simple and arbitrary with generators
$t_a$, $a = 1,2,...,N$. Let M$^4$ be
a differentiable but not necessarilly contractible base manifold. Thus, a
fibre bundle with $G$ as structure group (principal fibre bundle)
can be defined, assuming that the fibre proyection, the transition 
functions,
etc. are given.
A one-form connection arise with components
$A_\mu =A_\mu ^at_a(\mu =0,1,2,3)$, and transforms under the gauge group
in a usual manner
\begin{equation}
A_\mu ^{\prime }=UA_\mu U^{-1}+U\partial_\mu U^{-1}\,,\, U\in G. \label{eq1}
\end{equation}
Next, gauge invariant maps from $G$ into reals numbers are considered.
There are many ways to construct such maps.
For example, we can perform gauge invariant combinations on functional 
traces
in powers of $\Lambda _\mu (x)\equiv A_\mu (x)-\stackrel{\_}{A_\mu 
}(x)$, where
$\stackrel{\_}{A_\mu \,}$ is the background connection \cite{DTN} 
(arbitrary element
of the fibre). We will call this type of maps a Local Maps (LM).

On the other hand, we can considerate Non Local Maps (NLM) that involve
the Wilson functional or ''Wilson Loop'', which is defined as the 
holonomy trace:
$W(c) = trH(c)\equiv trP\exp \left( i\oint_cA\right)$, with
$c$ an element of the Group of Loops \cite{GT} inscribed in 3+1 space-time.

Given the above definitions we present two primary
considerations which will constitute a possible gauge formulation
for gravity:

1.- Assuming the existence of maps from $G$ onto reals numbers $R1$,
the metric tensor in 3+1 dimension is realized with
functionals that are gauge invariant under eq.(1) :
\begin{equation}
L.M.:\, g_{\mu \nu } (x)
\equiv S_{\mu \nu }\left( A(x) -\stackrel{\_}{A} (x)\right) \in R1 \,\,, 
\label{eq2}
\end{equation}
or
\begin{equation}
N.L.M.:\, g_{\mu \nu } (x)\equiv \, S_{\mu \nu }\left( A(x)
-\stackrel{\_}{A} (x),W(c)\right) \in R1 \,\,,\label{eq3}
\end{equation}

For any type of formulation, these relations will be written in the form:
$g_{\mu \nu }\left( x\right) \equiv S_{\mu \nu }\left( 
A\left(x\right)\right)$,
where $S_{\mu \nu}$ is some functional of $A\left(x\right)$.

2.- The lagrangian density is
\begin{equation}
L_G=-\frac 1{4k}\sqrt{-S\left(A\right) }F_{\mu \nu}^a(A)F^{a\mu \nu }(A) 
\,\,, \label{eq4}
\end{equation}
where $F_{\mu \nu }^a(A)=A_{\nu ;\mu }^a-A_{\mu ;\nu }^a+C^{abc}A_\mu
^bA_\nu ^c$ is the Yang-Mills curvature, $C^{abc}$ the completely 
antisymmetric
structure constant and $k$ a real constant.
We will assume that
eq.(4) is a Lagrangian reformulation for Einstein theory, if the action
$I_G = \int d^4x L_G\left(A_\mu \left(x\right)\right)$
is equal to the Einstein one,  $I_R= 1/16\pi \int
d^4x\sqrt{-g\left(x\right)}
R\left(g_{\mu \nu} \left(x\right)\right)$ for a given gauge group $G$.
This is,
\begin{equation}
I_G=I_R \,\,.\label{eq5}
\end{equation}
In virtue of the arbitrariness integration region
we can say that:
\begin{equation}
-\frac 1{4k} F_{\mu \nu}^a(A(x))F^{a\mu \nu}(A(x))
= \frac 1{16\pi} R\left(S_{\mu \nu} \left(A(x)\right)\right)+ 
\Omega^{\mu}_{;\mu}(x)\,\,
,\label{eq6}
\end{equation}
where $\Omega^{\mu} (x)$ is a function of order
$r^{-n}$
with $n>2$, for $r\longrightarrow \infty$. This condition on
$\Omega^{\mu} (x)$ establishes that the constraint (6) (or
"non-holonomical" constraint, due to arbitrariness of $\Omega^{\mu} 
(x)$) selects the
physical field $A(x)$ for a given functionals $S_{\mu \nu}$ .

These considerations can be argued as follows. The first one
presents the connection $A_\mu $ as the fundamental field
on the way to enlarge the symmetries, yielding an invariant
formulation under General Transformation of Coordinates
and gauge transformations.
Further, the only propagated degrees of freedom in this formulation 
corresponds
to the field $A_\mu $. Thus, the number of local degree of freedom can be
fixed taking an adecuate gauge group $G$.

The discussion about which group $G$ can be choosen, in order to match 
degrees of freedom
between this gauge formulation and gravity, is still open.
The complete constraint system will be considered elsewhere.

The second consideration is connected with the dynamical aspect
of the gauge field. A simplest gauge invariant first order lagrangian 
density is proposed.
At the same time, a non-holonomical constraint on field $A_\mu $ is 
presented.

When a particular form of $S_{\mu \nu}$ is performed
(i.e., combinations in powers of traces
on $\Lambda _\mu \equiv A_\mu -\stackrel{\_}{A_\mu}$, etc.),
the Yang-Mills like equations (see eq.(10)) are been prepared to conduce
a certain solution, but this $S_{\mu \nu}$ must be consistent with 
constraint (6).

\section{Dynamics}

Here, a variational analysis on the total action $I = I_G + I_M$ is 
performed.
$I_M$ is the matter fields action with variation
$\delta I_M=-1/2\int d^4x\sqrt{-g}T_M^{\alpha \beta }\delta g_{\alpha 
\beta }$.
Then, thinking in matter fields as external ones, arbitrary variations 
on $A_\mu $
yields
\begin{equation}
0=\int d^4x\sqrt{-S}\left[ -\frac 12\right( T_M^{\alpha \beta }+T_F^{\alpha
\beta}\bigg) \delta S_{\alpha \beta }(A)
+\frac 1k(D_\mu F^{\mu \lambda }\big) ^b\delta A_\lambda^b \bigg] , 
\label{eq7}
\end{equation}
where $T_M^{\alpha \beta }$ is the matter stress tensor,
$T_F^{\alpha \beta }$ is the Yang-Mills stress tensor
defined by:
\begin{equation}
T_F^{\alpha \beta }=-\frac 1k\left( F_\sigma ^{a\alpha }F^{a\sigma \beta }-%
\frac{S^{\alpha \beta }}4F_{\mu \nu }^aF^{a\mu \nu }\right), \label{eq8}
\end{equation}
where $S^{\alpha \beta }$ satisfies $S^{\alpha \beta }S_{\alpha \mu } 
=\delta _\mu ^\beta $
and $D_\mu $ is a covariant derivative under gauge transformations and
Coordinates transformations defined by
\begin{equation}
\left( D_\mu F^{\mu \nu }\right) ^b\equiv F_{;\mu }^{b\mu \lambda
}+C^{bac}A_\mu ^a F^{c \mu \lambda }\,\,. \label{eq9}
\end{equation}
Here, $F_{;\mu }^{b\mu \lambda }=\frac 1{\sqrt{-S}}\partial _\mu \left( 
\sqrt{%
-S}F^{b\mu \lambda }\right)$.

With an arbitrary
$\delta A_\mu $, for any formulation (LM or NLM), the field equations are:
\begin{equation}
\left( D_\mu F^{\mu \lambda }\right) ^b=\frac k2\left( T_M^{\alpha \beta
}+T_F^{\alpha \beta }\right) M_{b\, \alpha \beta }^\lambda \,\,, 
\label{eq10}
\end{equation}
where the object
\begin{equation}
M_{b\alpha \beta }^\lambda \equiv \frac {\partial S_{\alpha 
\beta}(A)}{\partial
A_\lambda^b} \,\,, \label{eq11}
\end{equation}
represents the "Jacobian" of the map that goes from $G$ to $R1$. In 
general it can
be observed that
the solutions of eq.(10) will depend on which prescription we take for 
the functional
$S_{\mu \nu}$ (satisfaying the constraint (6)).

In order to close this section we want to comment about the
relationship between the Yang-Mills like equations and Einstein ones for 
any gauge group $G$.
The gravity equations are:
\begin{equation}
R^{\alpha \beta }-\frac{g^{\alpha \beta }}2R+8\pi T_M^{\alpha \beta 
}=0\,\,, \label{eq12}
\end{equation}
with $R^{\alpha \beta }$ the Ricci tensor. Let us call $N^{\alpha \beta 
}=R^{\alpha \beta }-
\frac{g^{\alpha \beta }}2R+8\pi T_M^{\alpha \beta }$
a non null object,
when $R^{\alpha \beta }$ is not valued on Einstein equations (eq.(12)).
Then, equalizing the total action ($I = I_G + I_M$) with the obtained 
one from the
Einstein theory ($I = I_R + I_M$), and taking into account an arbitrary 
variation on fields
$A_\lambda ^b$ with $\delta S_{\alpha \beta }=M_{b\, \alpha \beta 
}^\lambda \delta A_\lambda ^b$ ,
the following equations can be obtained:
\begin{equation}
\left( D_\mu F^{\mu \lambda }\right) ^b-\frac k2\left( T_M^{\alpha \beta
}+T_F^{\alpha \beta }\right) M_{b\, \alpha \beta }^\lambda =
-\frac k{16\pi }N^{\alpha \beta }M_b^\lambda \, _{\alpha \beta }\,\, . 
\label{eq13}
\end{equation}

If eq.(12) (in other words, $N^{\alpha \beta }=0$) is introduced in eq.(13),
the dynamical equations (10) arise. However, in general would be 
possible to find
solutions for Yang-Mills like equations for a given functional
$S_{\mu \nu }(A(x))$ does not corresponds to general relativity solutions
(i.e., $S_{\mu \nu }(A(x)) \neq g_{\mu \nu }(x)$). This means in general
that the space of solutions of Yang-Mills like equations would contains the
Einstein's one.

\section{Non trivial vacuum solution for the abelian theory}

In order to obtain vacuum ($T_M^{\alpha \beta }=0$) stationary
solutions for the abelian case we take $G = U(1)\times 
\stackrel{N}{...}\times U(1)$.
In other words, we have $N$ generators that satisfy a Lie Algebra with 
$C^{abc}=0$
for all $a,b,c = 1,2,...N$. Moreover, the physical system consists in a 
compact spherical
symmetric stationary object. This fact allows to assume a stationary 
connection
of the type $A_\mu ^a =A_\mu ^a(r)$.

Thus, out of the compact object, the field equations looks as:
\begin{equation}
F_{;\mu }^{b\mu \lambda }=\frac k2T_F^{\alpha \beta }M_{b\, \alpha \beta 
}^\lambda \,\,.\label{eq14}
\end{equation}
If one want to solve these equations, we can still give a special form 
to the connection and metric.
Thinking in $A_\mu ^a\left( r\right)$, we take the electrostatic ansatz:
\begin{equation}
A_0^a\neq 0\,\,,\label{eq15}
\end{equation}
\begin{equation}
A_k^a=0\,\,\,\,.\label{eq16}
\end{equation}
For the metric we choose a Schwarzschild form
(diagonal and non time dependent):
\begin{equation}
diag(g_{00}\left( A_0^b\right) ,-1/g_{00}\left( 
A_0^b\right),r2,r^2sen2\theta ) ,\label{eq17}
\end{equation}
that will give an ansatz for the form of functional $S_{\mu \nu}(A(x))$.

Equations (15), (16) and (17) in (14) give
\begin{equation}
\frac{d^2A_0^a}{dr2}+\frac 2r\frac{dA_0^a}{dr}=0 \,\,\,\,,\label{eq18}
\end{equation}
\begin{equation}
A_k^a=0 , \label{eq19}
\end{equation}
and the solution is
\begin{equation}
A_0^a=-n^a+\frac{b^a}r\,\,,\label{eq20}
\end{equation}
\begin{equation}
A_k^a=0\,\,,\label{eq21}
\end{equation}
with constants $n^a$ and $b^a$. From eq.(6) is easy to probe that
$\Omega ^1 (x) \sim O(1/{r^{3}})$. This shows that the Schwarzschild ansatz
(see eq.(17)) is satisfactory.

An interesting case of eqs.(20) and (21)
occurs if we put $b^a=n^a 2m$, with the Schwarzschild radius $2m$:
\begin{equation}
A_\mu ^a=n^ag_{0\mu }\,\,,\label{eq22}
\end{equation}
pointing that $A_\mu ^a$ is proportional to the gravitational four-potential
of the Maxwell like formulation that C. M\"oller \cite{M} introduced in the
gravitational energy localization problem. In that reference, the author 
defines a
$U(1)$ potential:
\begin{equation}
A_\mu =g_{0\mu }\,\,, \label{eq23}
\end{equation}
covariant under the Space-Time Orthogonal transformations subgroup, 
given by:
\begin{equation}
x^{^{\prime }i}=f^i(x^j)\,\,, \label{eq24}
\end{equation}
\begin{equation}
x^{^{\prime }0}=x0. \label{eq25}
\end{equation}
Thus, relation (22) have the same subgroup of coordinate transformations.

Before ending this section, we want to explore the relation between the 
solution (22)
and the one obtained from other types of symmetries.
Is easy to probe for the Reissner-N\"ordstrom problem
that eq.(22) solves eq.(18) up an order $r^{-4}$ term. Obviously this is 
not a vacuum
problem because in this symmetry there is an
electrostatic charge. However, if $T_M^{\alpha \beta }$ is the stress tensor
associated to a rest charge, we have
\begin{equation}
0=\left( T_M^{\alpha \beta }+T_F^{\alpha \beta }\right) M_{b\, \alpha 
\beta }^\lambda
\,\,,\label{eq26}
\end{equation}
under the ansatz (15), (16) and (17). Thus, eq.(26) in (29)
throw an equation system equivalent to eqs.(18) and (19).

On the other hand, if one explore a non spherical symmetry (i.e., Kerr 
metric)
it can be shown that eq.(22) satisfies eq.(14)  up to
a term of order $r^{-5}$
(with an ansatz similar to (15) but with $A_1 ^a=A_2 ^a=0$ and $A_3 
^a=A_3 ^a(r,\theta )$).

All this says that the stationary solutions of the abelian formulation
($G= U(1)\times \stackrel{N}{...}\times U(1)$) for
Reissner-N\"ordstrom and Kerr symmetries, can be approached to eq.(22) 
at the
$r\rightarrow \infty $ limit. This asymptotic behaviour
is just the property that C. M\"oller \cite{M} needed
in his formulation in order to define the total energy in a satisfactory 
way.

\section{The Newtonian limit}

A required consistency condition of this formulation is that the Newtonian
limit must be recovered in the non relativistic weak field limit
from the Yang-Mills like equations.

In a low speed regime $\left( \left| v^i\right| =\left| dx^i/dx0\right| 
\ll 1\right)$
and a weak gravitational field, we take a Galilean Coordinate System
$x^\mu $ with a stationary metric $g_{\mu \nu }$ that defers up to a 
weak perturbation
$\left( \left| h_{\mu \nu }\right| \ll 1\right)$ from the Minkowski metric
$\left( \eta _{\mu \nu }\right)$, in other words:
\begin{equation}
g_{\mu \nu }=\eta _{\mu \nu }+h_{\mu \nu }\,\,, \label{eq27}
\end{equation}
with $g^{\mu \nu }=\eta ^{\mu \nu }-h^{\mu \nu }$.
Next is assumed a perfect fluid in which pressure and speed are
neglected in the Galilean System.
Only the $T_M^{00}$ component of material stress tensor (related with 
the mass density)
will be considered.

In order to complete the passage to the Newtonian limit from the
Yang-Mills like formulation we need to assume the next considerations.
First, it is required the linear behaviour of the gravitational field.
In this sense, the gauge group must be $G = U(1)\times 
\stackrel{N}{...}\times U(1)$.

Second, in the weak field regime, we can think that relation
(27) arises from a perturbation on the connection via eq.(2) or (3). In
others words, performing an infinitesimal variation $\delta A_{\mu }^b (x)$
around a fixed $A_{o \mu }^b$
($\Lambda _{o \mu} \equiv A_{o \mu } -\stackrel{\_}{A_{o \mu }}$
fixed too) with $S_{\mu \nu }(A_o )=\eta _{\mu \nu }$, we have
\begin{equation}
g_{\mu \nu } (x)=S_{\mu \nu }(A(x))=\eta _{\mu \nu }+
M_{b\mu \nu }^\lambda (A_o ) \delta A_{\lambda }^b (x)\,\,.\label{eq28}
\end{equation}

So then, with $M_{b\mu \nu }^\lambda (A_o)$
bounded, eq.(27) can be matched with eq.(28). Moreover, the
infinitesimal functions $h_{\mu \nu } (x)$
and $\delta A_{\mu }^b (x)$ have the same order.

On the other hand, $G = U(1)\times \stackrel{N}{...}\times U(1)$ in 
eq.(10) gives
\begin{equation}
F^{b\mu \lambda }_{;\mu} =\frac k2\left( T_M^{\alpha \beta }+T_F^{\alpha 
\beta
}\right) M_{b\, \alpha \beta }^\lambda \,\,,\label{eq29}
\end{equation}
and taking only first order contributions in $h_{\mu \nu }$
and $\delta A_{\mu}^b (x)$, the time component of the left hand side of 
eq.(29) yields
\begin{equation}
F^{bi0}_{;i}=-F_{i0,i}^b=-\nabla ^2A_0^b \,\,.\label{eq30}
\end{equation}

The Yang-Mills stress tensor
$\left( T_F^{\alpha \beta }\right)$ constitutes a quadratic order 
contribution
in $\delta A_{\mu }^b (x)$. With this, the time component of the right 
hand side of eq.(29) is
\begin{equation}
\frac k2\left( T_M^{\alpha \beta }+T_F^{\alpha \beta }\right) M_{b\,
\alpha \beta }^0 =\frac k2 M_{b 00}^0 (A_o ) T_M^{00} \,\,.\label{eq31}
\end{equation}

Joining eqs.(30) and (31) in (29) for $\lambda =0$, the next $N$ 
equations are obtained
\begin{equation}
\nabla ^2A_0^b=\alpha ^b T_M^{00} \,\,,\label{eq32}
\end{equation}
where $\alpha ^b = -\frac k2 M_{b 00}^0 (A_o )$. Expression (32) is the 
Laplace
equation for the Newtonian limit of the Yang-Mills like ones.

\section{Concluding remarks}

In this work, an initial idea about a possible scheme for a reformulation
of gravity theory in a similar way of a Yang-Mills theory at a classical 
level
was presented. This have been made thinking in two factibles gauge invariant
types of lagrangian formulations
(LM or NLM) which lead to consistent dynamical equations with the 
Newtonian limit.

A spherical symmetric stationary solution for the abelian case was obtained,
showing the relation with the Maxwell like four-potential of M\"oller.
It would be interesting to explore non abelian solutions for
stationary spherical symmetry where a
Bartnik-McKinnon \cite{BM} type ansatz can be used.

The fundamental problem oriented to
the quantisation bussines and related to the Dirac's canonical analysis
of constraints \cite{Dir} will be considerated elsewhere.

{\bf Acknowledgments }

The author thanks Profs. P\'{\i}o J. Arias and Luis Herrera
(Grupo de Campos y Part\'{\i}culas,
Departamento de F\'{\i}sica, Universidad Central de Venezuela) for 
observations.
Also thanks M. Botta Cantcheff (CBPF) and F. Brandt (Max Planck 
Institut) for references.

\end{document}